\documentclass[reprint,amssymb,superscriptaddress,aps,prl,floatfix]{revtex4-2}

\usepackage{amsmath}
\usepackage{array}
\usepackage[usenames,dvipsnames]{color}
\usepackage{lipsum}
\usepackage{tcolorbox}

\usepackage{printlen}

\usepackage{graphicx}
\usepackage{dcolumn}
\usepackage{bm}

\usepackage[utf8]{inputenc}   
\usepackage[T1]{fontenc}
\usepackage{lmodern}

\DeclareUnicodeCharacter{03B8}{\ensuremath{\theta}}   
\DeclareUnicodeCharacter{03C7}{\ensuremath{\chi}}     
\DeclareUnicodeCharacter{2212}{\ensuremath{-}}        
\DeclareUnicodeCharacter{2215}{/}                     
\DeclareUnicodeCharacter{2009}{\,}                    
\DeclareUnicodeCharacter{2011}{\nobreakdash-}         

\usepackage{hyperref}
\hypersetup{
  colorlinks=true,
  urlcolor=blue,
  citecolor=blue,
  linkcolor=black,
  ocgcolorlinks=false  
}

\makeatletter
\newcommand{\showfontsize}{The font size of the text is \f@size\,pt.}
\makeatother

\newcommand{\magenta}[1]{{\textcolor{black}{#1}}}
\begin{document}

\preprint{APS/123-QED}

\title{Holes in Sheets:\\Double-Threshold Rupture of Draining Liquid Films}

\author{Ayush K. Dixit}
\email{a.k.dixit@utwente.nl}
\affiliation{
	Physics of Fluids Department, Max Planck Center Twente for Complex Fluid Dynamics, and J. M. Burgers Center for Fluid Dynamics, University of Twente, P.O. Box 217, 7500AE Enschede, Netherlands
}

\author{Chunheng Zhao}
\email{czhao000@citymail.cuny.edu}
\affiliation{
	Department of Mechanical Engineering, City College of New York, New York, New York 10031, USA
}

\author{St\'ephane Zaleski}
\email{stephane.zaleski@sorbonne-universite.fr}
\affiliation{
	Sorbonne Universit\'e and CNRS, UMR 7190, Institut Jean Le Rond $\partial$'Alembert, 75005 Paris, France
}
\affiliation{
	Institut Universitaire de France, UMR 7190, Institut Jean Le Rond $\partial$'Alembert, 75005 Paris, France
}

\author{Detlef Lohse}
\email{d.lohse@utwente.nl}
\affiliation{
	Physics of Fluids Department, Max Planck Center Twente for Complex Fluid Dynamics, and J. M. Burgers Center for Fluid Dynamics, University of Twente, P.O. Box 217, 7500AE Enschede, Netherlands
}
\affiliation{
	Max Planck Institute for Dynamics and Self-Organisation, Am Fassberg 17, 37077 G{\"o}ttingen, Germany
}

\author{Vatsal Sanjay}
\email{vatsal.sanjay@comphy-lab.org}
\affiliation{CoMPhy Lab, Department of Physics, Durham University, Science Laboratories, South Road, Durham DH1 3LE, United Kingdom}
\affiliation{
	Physics of Fluids Department, Max Planck Center Twente for Complex Fluid Dynamics, and J. M. Burgers Center for Fluid Dynamics, University of Twente, P.O. Box 217, 7500AE Enschede, Netherlands
}

\date{\today}

\begin{abstract}

Classical rupture is attributed to molecular (van der Waals) forces acting at nanometric
thicknesses. Nonetheless, micron-thick liquid sheets routinely perforate far above the
scale where these molecular forces act, yet the mechanism that selects opening versus
healing has remained unclear. Using direct numerical simulations of a draining sheet
with an entrained air bubble (cavity), we show that irreversible rupture occurs only when a
deterministic double-threshold is crossed: (i) the outward driving (from airflow or inertia)
is strong enough and (ii) the cavity is distorted enough. If either condition
falls short, surface tension heals the cavity and the sheet reseals. The
time for this process is set by the balance between inertia and viscosity -- fast for inertia-dominated sheets
and slower for viscous ones. This double-threshold mechanism explains why micrometer-thick
films perforate and offers practical control options -- driving strength and defect
geometry --  for predicting and controlling breakup in spray formation processes, wave breaking, and respiratory films.

\end{abstract}

\maketitle

\begin{figure}
	\includegraphics[width=\columnwidth]{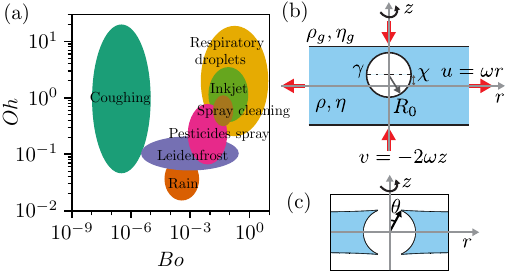}
  \caption{\label{fig:schematic} (a) Liquid sheets characterized by control parameters
  $Oh$ and $Bo$ are relevant in several phenomena across the entire parameter space. (b)
  Schematic side view of a liquid sheet that is radially draining and, subsequently, thinning
  along the axial direction. The flow directions of the liquid are indicated in red. A
bubble with radius $R_0$ is placed axisymmetrically but offset axially by a distance
$\chi$. (c) When additional physical factors delay the nucleation of the hole, the initial conditions
used in simulations are characterized by the polar angle $\theta$ made at the cavity edge.}
\end{figure}

The rupture of thinning liquid sheets is a ubiquitous route to fluid fragmentation, 
underlying phenomena from respiratory aerosolization to agricultural sprays and wave breaking \cite{deike2022mass}.
During the recent COVID-19 pandemic \cite{ciotti2020covid}, the public became acutely
aware of how violent expiratory events such as coughing and sneezing produce virus-laden
droplets. 
High-speed visualization of real-life coughs \cite{scharfman2016visualization,
bourouiba2021fluid, pohlker2023respiratory}, and experiments with mechanical cough devices
\cite{kant2023bag,li2025viscoelasticity} show that a mucosalivary liquid layer in the
respiratory tract can be sheared into a thin bag-like sheet that drains and thins rapidly.
Eventually, holes appear spontaneously in these thinning sheets, which expand and cause
the sheet to break apart, resulting in a cloud of droplets. 
Similar sheet rupture and drop-formation processes occur in a variety of contexts, 
including ocean wave breaking \cite{deike2022mass}, agricultural spray dispersal 
\cite{makhnenko2021review},
rain-induced aerosols \cite{villermaux2009single}, spray cleaning
\cite{josserand2016drop}, and drop impact on liquid pools
\cite{aljedaani2018experiments,lhuissier2013drop} or even solids \cite{kim2020raindrop}
(fig.~\ref{fig:schematic}a). 
Quantifying how such droplets are generated is crucial for
accurate risk assessments of airborne disease transmission \cite{bourouiba2021fluid},
optimizing pesticide applications \cite{makhnenko2021review}, and quantifying sea spray
aerosol generation \cite{de2011production}.

A key step in all of these processes is the nucleation of holes in the thinning sheet,
which sets the eventual droplet size distribution \cite{dombrowski1954photographic,
taylor1959dynamics}. 
However, despite its ubiquity, the physical origin of these holes remains
debated \cite{lohse2020double} -- especially for micrometer-thick sheets, where molecular 
mechanisms (van der Waals forces or thermal fluctuations) are too weak to explain rupture
\cite{neel2018spontaneous}.  Consistent with this view, experiments show that breakup
typically occurs once sheets thin to micron scales and that holes nucleate at internal
defects -- entrained bubbles or oil droplets -- within the sheet \cite{opfer2014droplet,
kant2023bag, jackiw2021aerodynamic, stumpf2023drop, aljedaani2018experiments,
thoroddsen2006crown}. In particular, bubbles in so-called Savart water sheets produced by a jet
impinging on a disk have been observed to perforate the sheet
\cite{lhuissier2013effervescent, gong2022effect}. Yet the pathway by which a bubble or
other impurity triggers hole formation—particularly under realistic sheet-drainage
conditions—remains unresolved because the nucleation events are rapid and difficult to
control and visualize.

In this Letter, we show that an air bubble entrained in a draining liquid sheet can
nucleate a hole long before molecular-scale forces become relevant. Axisymmetric direct
numerical simulations performed with the Navier-Stokes and combined level-set-volume-of-fluid solver \textsc{Basilisk C}
\cite{basilliskpopinet, supplMaterial, coderepository}\nocite{coderepository,tryggvason2011direct,
  brackbill1992continuum, sussman2000coupled, saini2025implementation,
popinet2015quadtree, popinet2009accurate, popinet2018numerical, duchemin2002jet,
deike2018dynamics, bouwhuis2012maximal} isolate this impurity-driven pathway under
controlled conditions while retaining realistic rim dynamics. To focus on the essential
physics, we replace the inflating bag geometry of air-blast breakup by an initially flat
sheet of thickness $h_0$ that drains radially with $u(r)=\omega r$
(fig.~\ref{fig:schematic}b),  \magenta{resulting in sheet-thickness far away from the bubble as $h = h_0 e^{-2\omega t}$}, a template that captures both external-airflow forcing in bag
breakup \cite{jalaal2012fragmentation} and inertia-driven drainage after drop impact \cite{sanjay2025unifying}. 
The dynamics are governed by the
Ohnesorge number $Oh=\eta/\sqrt{\rho\gamma R_0}$, the Bond number $Bo=\rho\omega^2
R_0^3/\gamma$, and the non-dimensionlized offset $\chi/R_0$ of the bubble's center from the centerlines of the flow (fig.~\ref{fig:schematic}b). Here $\eta$ and $\rho$ are the liquid
viscosity and density, $\gamma$ is the surface tension, and $R_0$ the bubble radius. The
driving acceleration $\omega^2 R_0$ acts as an effective radial gravity. In addition, the initial cavity distortion is characterized by an open angle $\theta$ (fig.~\ref{fig:schematic}c), see supplementary material \cite{supplMaterial}\nocite{coderepository,tryggvason2011direct,
	brackbill1992continuum, sussman2000coupled, saini2025implementation,
	popinet2015quadtree, popinet2009accurate, popinet2018numerical, duchemin2002jet,
	deike2018dynamics, bouwhuis2012maximal} for details on drainage mechanism leading to this geometry. 

Our two key results are as follows: First, the entrained bubble grows a through-cavity (a cavity that spans the entire film thickness, forming a continuous opening across the sheet) that overcomes surface tension and nucleates a hole at micrometer-scale thicknesses, explaining
rupture without invoking nanoscale physics. Second, the post-nucleation fate is
deterministic and requires crossing a double-threshold for irreversible opening: the outward
drainage must exceed a critical $Bo_c(Oh)$ and the initial cavity distortion -- characterized
by an opening angle $\theta$ -- must exceed a critical $\theta_c(Oh)$. Otherwise, rims
collide and the sheet heals. The opening-healing boundary collapses onto simple asymptotic
scalings: $Bo_c=\mathcal{O}(1)$ for $Oh\ll1$ and $Bo_c\sim Oh^{-2}$ for $Oh\gg1$. These
results rationalize the observed perforation of micron-thick sheets by internal defects
and provide a predictive framework for impurity-induced breakup in natural and engineered
fragmentation flows.

\begin{figure*}
	\includegraphics{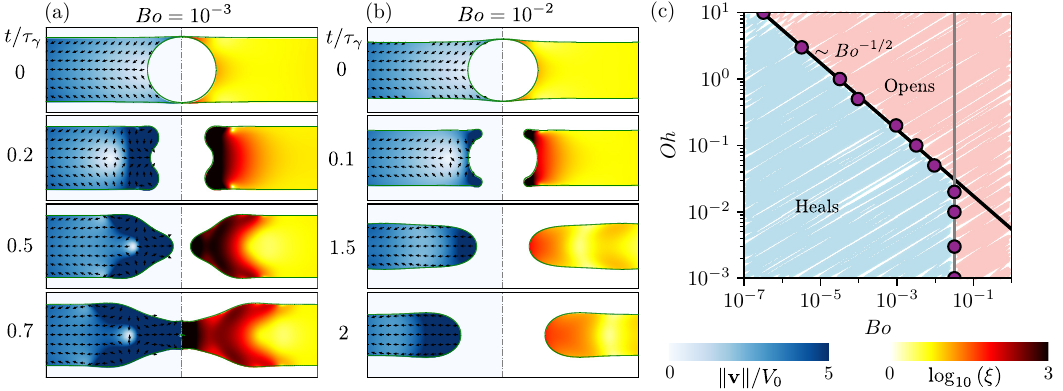}
  \caption{\label{fig:driving} As the liquid sheet drains radially and thins axially, the
  interfaces of the bubble and sheet merge to create a cavity. The time instants are
shown for $Oh = 0.1$, (a) and at small $Bo = 10^{-3}$ where the sheet heals, while (b) at
larger $Bo=10^{-2}$, the sheet opens up. The left panel depicts velocity magnitude
$\|\boldsymbol{v}\|/V_0$, where $V_0 = \sqrt{Bo \gamma/\rho R_0}$, and the black arrows
depict the velocity direction. The right panel illustrates viscous dissipation $\xi = 2
Bo \left(\boldsymbol{\mathcal{D}:\mathcal{D}}\right)$, where $\boldsymbol{\mathcal{D}} =
\left(\boldsymbol{\nabla u + \nabla u}^T\right)/2$ is the symmetric part of velocity
gradient tensor. (c) \magenta{Effect of driving and viscosity on sheet rupture}. The regime map in the log-log parameter space of $Oh-Bo$. The transition lines at small $Oh$ are shown by a constant $Bo$ line in gray, while, at
large $Oh$, the transition is indicated by a black line with scaling $Oh \sim Bo^{-1/2}$.
}
\end{figure*}

{\it Phenomenology} -- 
We first examine sheet-centered bubbles ($\chi/R_0 = 0$) across the
$(Oh,Bo)$ space. As the sheet thins, its top and bottom interfaces meet at the bubble
poles and a through-cavity opens. \magenta{In the numerical simulations, the film between the bubble and sheet interface ruptures when the local thickness reaches the smallest grid size, chosen to correspond to approximately 10 nm in physical units to mimic van der Waals-driven rupture.} Capillary waves invert the cavity from convex to
concave, after which two outcomes are observed (fig.~\ref{fig:driving}a,b): either the rims
retract and the hole expands, breaking the sheet, or capillarity drives rim collision and
the sheet heals. A systematic sweep in $(Oh,Bo)$ reveals a sharp transition curve
(fig.~\ref{fig:driving}c): strong radial driving (large $Bo$) yields opening, whereas
weaker driving permits healing unless $Bo$ exceeds a viscosity-dependent threshold. The
location and slope of this boundary reflect the competition between capillary-driven rim
closure and drainage-driven advection, anticipating the timescale analysis that follows.
We additionally show that delayed nucleation--representing chemical/thermal
heterogeneity--can be encoded by a larger initial cavity characterized by an opening angle
$\theta$, which shifts the transition and thereby introduces geometry as a second way of control.

\begin{figure*}
 	\includegraphics{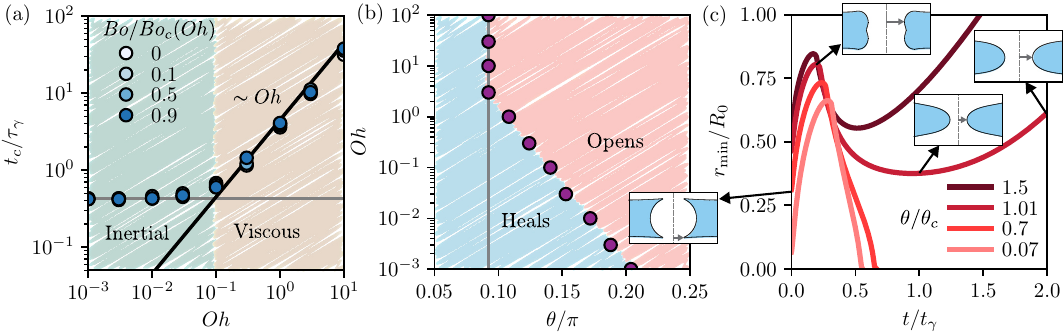}
  \caption{ (a) After the bubble cavity opens, in some cases, capillarity manages to heal
    the rims, and the time taken is referred to as the collision time $t_c$. Here,
    $t_c/\tau_\gamma$ is plotted against $Oh$ at several $Bo$, where $\tau_\gamma =
    \sqrt{\rho R_0^3/\gamma}$. At small $Oh$, the gray line shows the scaling of the
    inertio-capillary timescale, while the black line shows the visco-capillary time scale
    at large $Oh$; both scalings seem to be consistent with simulation results. (b) The parameter space of $Oh-\theta$ highlighting the healing and opening regimes with different colors. The gray line shows the transition
    observed at $\theta = 0.09\pi$ at large $Oh$, while the individual data points are
    also denoted. (c) The evolution of minimum tip radius $r_{\text{min}}$ at several
    initial distortions $\theta$ for $Oh = 0.1$ without external driving. For smaller
  distortions, $r_{\text{min}}$ decays to zero, whereas for larger distortions, sheets eventually 
open irreversibly, consistent with Taylor-Culick retractions. At several instances, the insets depict sheet profiles and highlight $r_{\text{min}}$ with gray arrows.
 	\label{fig:shape}}
 \end{figure*}

{\it Relevant timescales set the healing to opening transition} -- 
After the bubble's and sheet's interfaces merge and create an initial cavity through the sheet, we
measure the time it takes for the two opposing rims to collide at the sheet's central
axis; we term this the ``collision time'' $t_c$. Physically, $t_c$ represents the healing
time when the sheet successfully reseals. Fig.~\ref{fig:shape}(a) shows $t_c$ (normalized
by the inertio-capillary timescale $\tau_\gamma = \sqrt{\rho R_0^3/\gamma}$) as a function
of $Oh$ for several values of $Bo$ (including $Bo$ just below the opening threshold for
each $Oh$). At low viscosity ($Oh \ll 0.1$), we find that $t_c \sim \tau_\gamma$,
insensitive to $Oh$. At higher viscosity ($Oh \gg 0.1$), $t_c/\tau_\gamma \sim Oh$.

These trends can be understood by considering the dominant balance of forces resisting
hole closure. For small $Oh$ (inertia-dominated regime), viscous resistance is negligible
and inertial forces dominate. Balancing surface tension and inertia yields a
characteristic rim velocity $U_0 \sim \sqrt{\gamma/(\rho R_0)}$ and thus $t_c \sim
\tau_\gamma \equiv \sqrt{\rho R_0^3/\gamma}$. In contrast, for large $Oh$
(viscosity-dominated regime), balancing surface tension with viscous stress (of order
$\eta U_0/R_0$) gives $U_0 \sim \gamma/\eta$ and hence $t_c \sim \eta R_0/\gamma \sim
\tau_{\gamma} Oh$, a visco-capillary timescale. These scaling predictions agree well with
the simulation results in fig.~\ref{fig:shape}(a).

Notably, even when $Bo$ is very close to its critical value $Bo_c(Oh)$ (beyond which the
sheet would open), the measured healing time $t_c$ is essentially the same as for much
smaller $Bo$. Thus, while radial driving ($Bo$) determines whether the sheet ultimately
opens or heals, the healing dynamics themselves is primarily controlled by the liquid's
inertial-viscous balance ($Oh$).

We can predict the boundary between opening and healing regimes by comparing the
timescales for rim retraction versus radial sheet advection. For the sheet to open,
outward advection (driven by $\omega$, with timescale $t_{\text{adv}} \sim 1/\omega$) must
outpace capillary-driven rim closure (timescale $t_c$). At low viscosities (small $Oh$,
where $t_c \sim \tau_\gamma$), the criterion $t_c \sim t_{\text{adv}}$ leads to
$\sqrt{\rho R_0^3/\gamma} \sim 1/\omega$, yielding $Bo \sim 1$ in dimensionless terms. At
high viscosities (large $Oh$, where $t_c \sim \eta R_0/\gamma$), the condition $t_c \sim
1/\omega$ gives $Oh \sim Bo^{-1/2}$. These theoretical thresholds -- shown as the gray
(vertical) and black (diagonal) lines in fig.~\ref{fig:driving}(c) -- closely match the
transition observed in our simulations.

{\it What happens in the limiting case of no external driving ($Bo=0$)?} -- 
In the absence of radial forcing ($Bo=0$), the liquid sheet inevitably heals for any
finite $Oh$ -- even extremely viscous sheets (large $Oh$) will eventually reseal, albeit
very slowly as viscosity prolongs the rim collision time. 
This follows from our thresholds: as $Bo \to 0$, achieving opening would require $Oh \sim
Bo^{-1/2} \to \infty$, so any finite $Oh$ lies in the healing regime. Our simulations
confirm this universal healing at $Bo=0$, consistent with the classic geometric energy
criterion of Taylor \& Michael \cite{taylor1973making}. Using a toroidal model for the
hole's rim \cite{energyCriterion}\nocite{ilton2016direct, goudarzi2023hole,
moriarty1993dynamic, sharma1990energetic}, they predicted that without external driving a
circular hole can continue to expand only if its radius exceeds a purely geometric
threshold. 
In particular, when the hole's outer radius $R$ \cite{torus} satisfies $R/h_0 > \pi/4$ (so that the
inner radius $R_h = R - h_0/2$ is about $0.29h_0$), the excess surface energy $\Delta \Xi$
becomes positive, signaling that the hole will grow rather than close. Holes smaller than
this critical size have $\Delta \Xi < 0$ and will heal. Notably, this threshold is
independent of viscosity. Thus, at $Bo=0$ no finite-$Oh$ case can open unless the initial
hole is above the critical $R/h_0$ -- indeed, the cavities in our simulations were
below this size, so they all heal.

In more realistic situations, additional physics can delay the initial rupture (for
example, due to chemical or thermal heterogeneity), allowing the sheet to drain longer and
thin more broadly before the hole forms. We represent such delays in our model by a larger
initial cavity distortion characterized by an increased polar angle $\theta$
(fig.~\ref{fig:schematic}c). 
Simulations at $Bo=0$ with varied $\theta$ reveal a clear geometry-controlled threshold: for
sufficiently large distortions (large $\theta$), the hole opens even without any driving,
whereas for smaller initial distortions the hole heals (fig.~\ref{fig:shape}b). 
This confirms the second threshold of the double-threshold framework: 
a sufficiently large initial cavity alone can trigger
sheet rupture in the absence of external forcing. 	Moreover, at very high viscosity (large Oh), the opening–healing transition occurs at a nearly constant opening angle $\theta \approx 0.09\pi$. Remarkably, this transition, when expressed in terms of $R_h/h_0$, gives the same result as the Taylor--Michael criterion $R_h/h_0 \approx 0.29$. However, as $Oh$ decreases
(inertia-dominated regime), this critical distortion $\theta_c$ rises, since
added inertia aids rim closure—requiring a larger initial opening to overcome healing. 
Remarkably, sheets that do open exhibit non-monotonic cavity growth (fig.~\ref{fig:shape}c). Surface tension first drives the cavity edges apart, causing radial cavity growth, but the induced inertia pulls the rim inward and temporarily reduces the cavity radius. However, the cavity remains large enough to overcome healing by this recoil and eventually opens with the Taylor--Culick mechanism.

\begin{figure*}
	\includegraphics{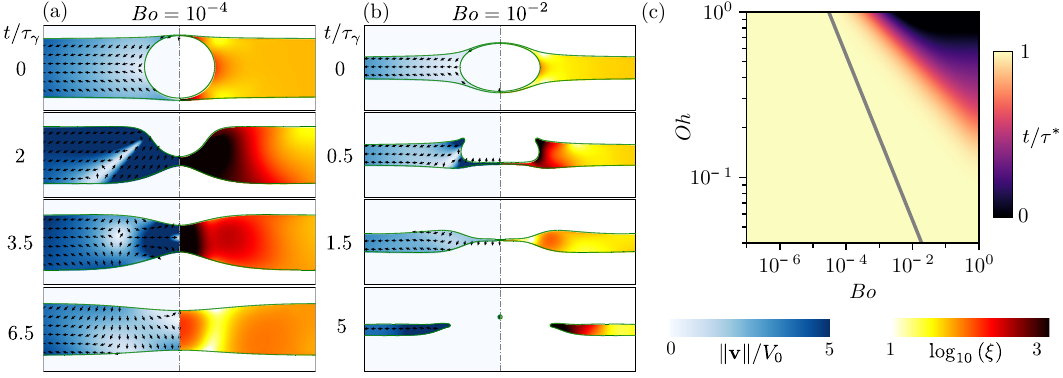}
  \caption{\label{fig:asymmetry} The draining liquid sheet of the bubble is placed
  asymmetrically with $\chi/R_0 = 0.1$. Time evolution has been shown for cases with
$Oh=1$, (a) $Bo=10^{-4}$, and (b) $Bo=10^{-2}$. In the former case, the liquid bridge at
the south pole replenishes and grows into a flat-shaped sheet due to capillary action, and
then the sheet keeps on thinning due to radial drainage. Meanwhile, in the latter case,
thinning due to radial drainage dominates before the bridge can be replenished by the
damped capillary waves at large $Oh=1$. The left panel depicts velocity magnitude
$\|\boldsymbol{v}\|/V_0$, where $V_0 = \sqrt{Bo \gamma/\rho R_0}$, the black arrows depict
the velocity direction. The right panel depicts viscous dissipation $\xi = 2 Bo
\left(\boldsymbol{\mathcal{D}:\mathcal{D}}\right)$, where $\boldsymbol{\mathcal{D}} =
\left(\boldsymbol{\nabla u + \nabla u}^T\right)/2$ is the symmetric part of velocity
gradient tensor. (c) \magenta{Effect of the asymmetric bubble position on sheet breakup}. Sheet breakup time (normalized by the no-bubble case
rupture time \magenta{$\tau^{*} = 3.5/\omega \equiv 3.5 \tau_\gamma/\sqrt{Bo}$, with the prefactor set by the grid cutoff}) in the $Oh-Bo$ parameter space for
$\chi/R_0 = 0.1$. \magenta{The plot is constructed from simulation results at 324 points arranged on a uniform logarithmic $18 \times 18$ grid in parameter space.} The gray line depicts the transition line observed for symmetric cases
($\chi/R_0 = 0$), as shown in fig.~\ref{fig:driving}. }
\end{figure*}

{\it Effect of off-center bubbles} --
Thus far we considered sheet-centered bubbles ($\chi/R_0 = 0$) with simultaneous holes at
both poles; in practice, an off-center bubble ($\chi/R_0 \neq 0$) breaks this symmetry.
For $\chi/R_0 > 0$, only the thinner pole ruptures initially, leaving a draining liquid
bridge at the opposite pole. If the offset is very small ($\chi \to 0$), that remaining
bridge is extremely thin and often ruptures almost immediately due to van der Waals forces
-- essentially the symmetric outcome.
For moderate asymmetry, however, the first opening launches a capillary wave that travels
toward the far pole and drives fluid into the intact bridge (fig.~\ref{fig:asymmetry}),
temporarily replenishing it and suppressing prompt hole nucleation there -- unlike the
symmetric case (fig.~\ref{fig:asymmetry}a). 

Nonetheless, even without an immediate second hole, the bubble's presence can
significantly shorten the sheet's lifetime, especially at high $Oh$. At low viscosity
(small $Oh$) and weak driving, the capillary wave significantly thickens the far-side liquid
bridge, so the sheet drains nearly as if no bubble were present
(fig.~\ref{fig:asymmetry}a). In this regime the bubble has minimal effect on the breakup
time. At higher viscosity (large $Oh$), however, capillary replenishment is strongly
damped; the bridge thins continuously under drainage and ruptures much earlier than it
would in a sheet without bubble (fig. \ref{fig:asymmetry}b).
Fig.~\ref{fig:asymmetry}(c) quantifies these trends by mapping the sheet breakup time
(normalized by the no-bubble value $\tau^{\ast}$) across the $Oh$-$Bo$ parameter space for ${\chi}$ = 0.1$R_0$.
Consistent with the above description, the bubble significantly hastens
breakup in the high-$Oh$, high-$Bo$ regime (dark regions), whereas at low $Oh$/low $Bo$
the breakup time is nearly unchanged from the bubble-free case (light regions). Comparing
to the symmetric configuration (gray transition line from fig.~\ref{fig:driving}), we see
that an off-center bubble is generally less effective at triggering rupture: with one
pole's bridge intact, the sheet is less prone to instantaneous rupture, and the parameter
region of strong bubble influence is reduced. In the limit of vanishing asymmetry, the
behavior converges to the symmetric case (points approaching the gray line in
fig.~\ref{fig:asymmetry}c). Further analysis of varying ${\chi}$ is provided in
\cite{supplMaterial}.

{\it Conclusion \& Outlook} --
We have shown that a micron-thick draining sheet pierced by a trapped bubble undergoes
irreversible rupture only if a double-threshold is exceeded: the driving (e.g., airflow or
inertia) must exceed a critical Bond number $Bo_c(Oh)$, and the initial cavity distortion
-- captured by a geometric opening angle $\theta$ -- must exceed a threshold
$\theta_c(Oh)$. If either threshold is unmet, capillarity heals the sheet on an inertio-
($Oh \ll 1$) or visco-capillary ($Oh \gg 1$) timescale. This double-threshold mechanism is
reminiscent of hole formation in drops impacting on rough surfaces, where a sufficiently
high inertia and a large surface roughness are both required for holes to nucleate
\cite{lohse2020double, kim2020raindrop, kolinski2014drops}. Similar double-thresholds govern other nonlinear
transitions in fluids: for example, turbulence in shear flow develops only when the
Reynolds number is high and the initial perturbation amplitude is large
\cite{avila2023transition}, and elastic turbulence in polymer solutions likewise demands a
large Weissenberg number together with strong disturbances \cite{samanta2013elasto}.
Importantly, an isolated flat liquid sheet in air is linearly stable as small distortions
decay and rupture only occurs through finite amplitude distortions \cite{eggers2011subtle,
  wang2015nonlinear, matsuuchi1974modulational, mehring1999nonlinear,
squire1953investigation}. This nonlinearity of the healing--opening transition is
reflected in our simulations, where small initial cavities decay while sufficiently large
ones rupture (fig.~\ref{fig:shape}c). However, unlike the subcritical double-threshold
mechanism behavior in transition to turbulence, our system is not subcritical and thus not hysteretic: once parameters
($Bo$, $\theta$, $Oh$) are set, the outcome (healing or opening) is uniquely determined.

A trapped bubble -- acting as a hydrophobic defect -- thus rationalizes why rupture occurs
far above molecular scales and provides predictive control knobs, namely driving strength
and defect geometry, relevant to bag-breakup sprays and respiratory films. 
\magenta{When the bubble is off-center, however, the sharp $Oh_c \sim Bo^{-1/2}$ transition 
gives way to a gradual crossover governed by the competing timescales of capillary-wave replenishment, 
viscous damping, and radial drainage -- physics that awaits a predictive scaling theory.} 
\magenta{In practice, small off-center bubbles at low $Oh$ and $Bo$ are unlikely
	to trigger rupture unless delayed nucleation (e.g., from chemical or thermal
	heterogeneity) enlarges the initial cavity sufficiently.}
Future studies could test whether analogous thresholds govern rupture triggered by solid particles, oil
droplets, or Marangoni-driven inhomogeneities. Chemical or thermal gradients may couple
nonlinearly with radial drainage and shift $Bo_c(Oh)$ and $\theta_c(Oh)$, extending this
double-threshold framework to a broader class of impurity-triggered, geometry-sensitive
instabilities.

\begin{acknowledgments} 
  {\it Acknowledgments:} We acknowledge the funding from the MIST consortium with project
  number P20-35 of the research programme Perspectief, which is (partly) financed by the
  Dutch Research Council (NWO). We also acknowledge the NWO-Canon grant FIP-II grant. This
  work was carried out on the national e-infrastructure of SURFsara, a subsidiary of SURF
  cooperation, the collaborative ICT organization for Dutch education and research. This
  work was sponsored by NWO - Domain Science for the use of supercomputer facilities. St\'ephane Zaleski and Chunheng Zhao were funded by the European Research Council (ERC) through the European Union's Horizon 2020 Research and Innovation Programme (Grant Agreement No. 883849 TRUFLOW).
\end{acknowledgments}


\bibliographystyle{prsty_withtitle}
\bibliography{holeySheets}

\begin{thebibliography}{10}
\bibitem{torus}
A torus is the surface of revolution obtained by rotating a circle about an
  axis lying in the same plane as the circle. The distance between the axis of
  revolution and the circle's center is referred to as the outer radius $R$,
  while the distance from the axis to the innermost edge of the surface is the
  inner radius $R_h$, given by $R_h = R - h_0/2$.
\end{thebibliography}

\end{document}


\preprint{APS/123-QED}

\title{Supplementary material: \\Holes in Sheets: Double-Threshold Rupture of Draining Liquid Films}

\author{Ayush K. Dixit}
\email{a.k.dixit@utwente.nl}
\affiliation{
	Physics of Fluids Department, Max Planck Center Twente for Complex Fluid Dynamics, and J. M. Burgers Center for Fluid Dynamics, University of Twente, P.O. Box 217, 7500AE Enschede, Netherlands
}

\author{Chunheng Zhao}
\email{czhao000@citymail.cuny.edu}
\affiliation{
	Department of Mechanical Engineering, City College of New York, New York, New York 10031, USA
}

\author{St\'ephane Zaleski}
\email{stephane.zaleski@sorbonne-universite.fr}
\affiliation{
	Sorbonne Universit\'e and CNRS, UMR 7190, Institut Jean Le Rond $\partial$'Alembert, 75005 Paris, France
}
\affiliation{
	Institut Universitaire de France, UMR 7190, Institut Jean Le Rond $\partial$'Alembert, 75005 Paris, France
}

\author{Detlef Lohse}
\email{d.lohse@utwente.nl}
\affiliation{
	Physics of Fluids Department, Max Planck Center Twente for Complex Fluid Dynamics, and J. M. Burgers Center for Fluid Dynamics, University of Twente, P.O. Box 217, 7500AE Enschede, Netherlands
}
\affiliation{
	Max Planck Institute for Dynamics and Self-Organisation, Am Fassberg 17, 37077 G{\"o}ttingen, Germany
}

\author{Vatsal Sanjay}
\email{vatsali.sanjay@comphy-lab.org}
\affiliation{CoMPhy Lab, Department of Physics, Durham University, Science Laboratories, South Road, Durham DH1 3LE, United Kingdom}
\affiliation{
	Physics of Fluids Department, Max Planck Center Twente for Complex Fluid Dynamics, and J. M. Burgers Center for Fluid Dynamics, University of Twente, P.O. Box 217, 7500AE Enschede, Netherlands
}

\date{\today}

\maketitle

\tableofcontents

\section{Governing equations}\label{sec:gonverning}

The governing dynamical equations have been solved using the free software Basilisk C  \citep{basilliskpopinet, popinet2015quadtree}. For all the quantities, length scales are normalized using the initial bubble radius, resulting in $\mathcal{L} = \tilde{\mathcal{L}}R_0$ as characteristic length, while the time is normalized using the inertiocapillary timescale $\tau_\gamma = \sqrt{\rho_s {R_0}^3/\gamma}$ giving $t = \tilde{t}\tau_{\gamma}$. These normalizations leads to an inertiocapillary velocity scale $u_{\gamma} = \sqrt{\gamma/ \rho_{s} R_0}$ for the velocity field $\boldsymbol{u} = \tilde{\boldsymbol{u}}u_\gamma$. Lastly, stresses are normalized using the Laplace pressure scale, $\boldsymbol{\sigma} = \tilde{\boldsymbol{\sigma}}\sigma_\gamma$, where $\sigma_\gamma = \gamma/R_0$. The governing mass and momentum conservation equations for the liquid phase in dimensionless form read 
\begin{align}
	\label{eq:massconserve}
	\boldsymbol{\nabla\cdot u}=0 \, \text{and}
\end{align}

\begin{equation}
	\frac{\partial \boldsymbol{u}}{\partial t} + \boldsymbol{\nabla\cdot} \left(\boldsymbol{u}\boldsymbol{u}\right) =  -\boldsymbol{\nabla}p + 2\ Oh\boldsymbol{\nabla\cdot}\boldsymbol{\mathcal{D}}, 
	\label{eq:momconserve}
\end{equation}

\noindent where $\boldsymbol{\mathcal{D}} = \left(\boldsymbol{\nabla u} + \left( \boldsymbol{ \nabla u} \right) ^T \right)/2$ represents the symmetric part of the velocity gradient tensor--equal to half of the rate-of-strain tensor, and $p$ is the pressure field. 

\section{Numerical method}\label{sec:methods}
We build upon and employ the open-source software Basilisk C \citep{basilliskpopinet, popinet2015quadtree} to simulate the draining bubbly sheet. The utilized code is shared openly in our repository \cite{coderepository}. The governing equations are solved using the one-fluid approximation \citep{tryggvason2011direct}, with surface tension incorporated as singular body force at the liquid-gas interface \cite{brackbill1992continuum}. To account for the gas phase, we maintain a constant Ohnesorge number based on air viscosity, i.e., $Oh_a = 2 \times 10^{-5}$, and a constant density ratio $\rho_g/\rho = 0.001$. The liquid-gas interface is tracked using the Coupled Level Set volume of fluid (CLSVoF) method. CLSVoF combines some of the advantages of VoF and Level Set (LS) method \cite{sussman2000coupled, basilliskpopinet, saini2025implementation}; this method is mass conserving, which is the advantage of VoF, while the curvature is estimated using the signed distance function, which is the advantage of LS. The interface is tracked using the VoF method governed by the advection equation

\begin{align}
	\frac{\partial \Psi }{\partial t} + \boldsymbol{\nabla\cdot}\left( \Psi \boldsymbol{u}\right) = 0. 
	\label{eq:volfracconserve}
\end{align}

\noindent where $\Psi$ represents the VoF color function. We implement a geometric VoF approach reconstructing the interface at each time step, while the signed distance field $\mathsf{d}$ is advected as a tracer, which is obtained from VoF reconstruction of the interface. The distance is combined with the existing distance using a small weight. Finally,  $\mathsf{d}$ is used to estimate the surface tension forces acting as singular forces \cite{popinet2009accurate, brackbill1992continuum}. The explicit treatment of surface tension imposes a time step constraint based on the smallest capillary wave oscillation period \citep{popinet2009accurate}.


The axisymmetric liquid sheet is simulated in the computational domain spanning $6R_0 \times 6R_0$. A bubble is placed centered on the axis and off the distance $\chi$ from the center of the sheet. In all simulations, the film initially has a thickness of $6R_0$, while the radius of the entrapped bubble is $R_0$. The subsequent dynamics is independent of this initial thickness because the interfacial velocity, $v = -2\omega z$, depends solely on the instantaneous film thickness and not on its initial value. Our simulations consistently confirm that the final outcome remains unaffected by the initial film thickness, provided it is sufficiently large. The chosen initial thickness should be sufficiently large to ensure that the flow-driven lubrication-layer formation between the bubble and the sheet interfaces is not hindered. In our simulations, sheet drainage, caused by effective radial acceleration, is enforced through the boundary conditions, where the initial velocity condition matches the velocity of the draining sheet without a bubble. Consequently, a few initial timesteps are required for the flow to establish accurate velocities in the lubrication region. If the simulation were initialized with a film thickness close to $h_0 \approx 2R_0$, the sheet would rupture too rapidly for the flow to fully develop near the bubble, potentially introducing small deviations in the results. To avoid this, the initial film thickness is chosen sufficiently large to ensure consistency and eliminate such artifacts. Rupture of the film between the bubble–sheet interface numerically occurs when the local film thickness equals the smallest grid size. To mimic van der Waals–driven rupture, this minimum grid size is chosen such that it corresponds to approximately 10 nm in physical units. At the domain boundaries, velocity conditions are imposed, i.e., $u=\omega \left(6 R_0\right)$ and $v= - 2 \omega \left(6R_0\right)$, while the pressure gradients are set to zero for both liquid and gas phases. To simulate cases without any external driving, a very narrow region is initially removed at the contact between the sheet of thickness $2R_0$ and bubble interfaces. The domain is discretized using quadtree grids with adaptive mesh refinement (AMR) \citep{popinet2009accurate}. Error tolerances for the VoF color function, curvature, and velocity are set to $10^{-3}$, $10^{-6}$, and $10^{-3}$, respectively.

\section{Characterize and estimate the initial cavity shape influenced by additional physical inhomogeneities}\label{sec:shape}

\begin{figure}[h!]
	\includegraphics{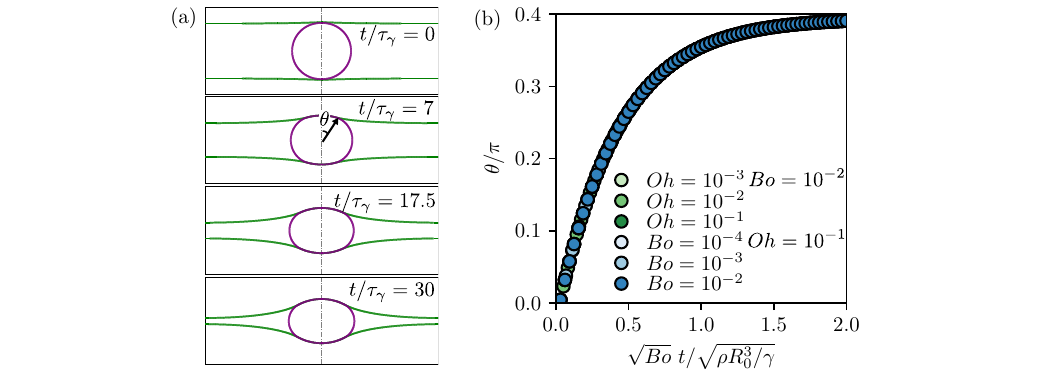}
	\caption{\label{fig:ab_time}  (a) The time evolution of the draining sheet when the bubble-sheet interfaces do not coalesce at $Oh=0.1$, and $Bo=10^{-3}$. The green and magenta depict the sheet and bubble interfaces, respectively. The shape is characterized by the angle $\theta$, which is the circumferential angle covered by the overlapping bubble-sheet interfaces. (b) The evolution of $\theta$ against $ \sqrt{Bo} \ t_r/\sqrt{\rho R_0^3/\gamma}$ for various $Bo$ and $Oh$ higlights that the shape evolves independently of $Bo$ and $Oh$ and is solely determined by $ \sqrt{Bo} \ t_r/\sqrt{\rho R_0^3/\gamma}$. 
	}
\end{figure}

As discussed in the main text, in realistic scenarios, chemical or thermal inhomogeneities can delay initial hole nucleation. This delay leads to the formation of a wider thin film region near the poles, which eventually bursts to create a larger cavity. To simulate the initial shape of the sheet with a cavity, we conduct simulations where the interfaces of the bubble and sheet do not coalesce. When the sheet interfaces come into contact with bubble interfaces, instead of merging to create the cavity, sheet interfaces persist on bubble interfaces without coalescence. This is accomplished numerically by introducing different tracers for bubble and surrounding air. The time evolution for non-coalescing simulations is shown in fig. \ref{fig:ab_time}. At any time $t_r$, the thin bridge region is manually ruptured to obtain the cavities characterized by $\theta$ -- large $t_r$ corresponds to wider cavities, and thus, large $\theta$. Therefore, for simulations to assess the effect of different initial cavities, the initial condition is taken as the profile at time $t_r$ after removing the thin film cap. This approach has been widely adopted in numerous studies related to floating bubbles \cite{duchemin2002jet, deike2018dynamics}. 

The initial cavity profile is characterized by the angle $\theta (Bo, t_r)$, which is obtained by selecting the parameters $Bo$, and $t_r$. However, remarkably, the shape is observed to be the same, regardless of $Bo$, and $t_r$ as long as $\sqrt{Bo} \times t_r$ is same. Thus, the initial cavity shape $\theta(\sqrt{Bo} \ t_r)$ is observed to be solely determined by $\sqrt{Bo} \times t_r$ for $Bo < 10^{-2}$ which includes almost all the realistic scenarios. For higher $Bo > 10^{-2}$, unlike small $Bo$ cases, the minimum bubble-sheet thickness is no longer at the poles; instead, a small liquid volume is sandwiched at the poles, and the minimum thickness is observed off the poles. This observation poses a resemblance with the drop impact phenomenon \cite{bouwhuis2012maximal}, where for small Stokes number $St$, no air bubble is entrapped below the droplet. Also, the minimum thickness of the drop-plate interface is at the poles--similar behavior is observed here for $Bo < 10^{-2}$. In contrast, for larger $St$, air bubbles are entrapped, and the minimum thickness is observed away from the poles, which is similar to behavior here at larger $Bo > 10^{-2}$. 

 \section{Effects of bubble asymmetry $\chi/R_0$}\label{sec:asymmetry}
\begin{figure}[h!]
	\includegraphics{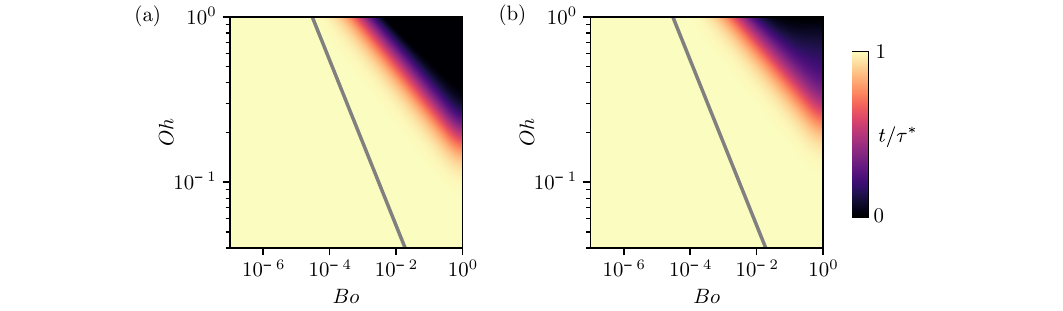}
	\caption{\label{fig:OhBo-chi}  Sheet breakup time (normalized by the no-bubble case rupture time \magenta{$\tau^{*} = 3.5/\omega \equiv 3.5 \tau_\gamma/\sqrt{Bo}$, with the prefactor set by the grid cutoff}) in the $Oh-Bo$ parameter space, at (a) $\chi/R_0 = 0.05$, and (b) $\chi/R_0 = 0.15$. \magenta{The plots are constructed from simulation results at 324 points arranged on a uniform logarithmic $18 \times 18$ grid in parameter space.} The gray line depicts the transition observed for symmetric cases ($\chi/R_0 = 0$). 
	}
\end{figure}

When the bubbles are placed asymmetrically in axial direction ($\chi/R_0 > 0$), the sheet no longer ruptures immediately, and the presence of the intact liquid bridge reduces the tendency of the sheet to break compared to the corresponding symmetric case, which is the least stable one. 
In the asymmetric cases illustrated in fig. \ref{fig:OhBo-chi}, the darker zones depict the regions where the effect of the initial bubble on bubble breakup is still prominent. These regions are narrower compared to the opening regime of symmetric cases at the right side of the gray transition line, bolstering the fact that the addition of asymmetricity reduces the tendency of sheets to break. As $\chi/R_0$ increases, the effect of the asymmetry becomes even more pronounced. That is, the tendency of the sheets to break reduces even further. This aspect is also evident in fig. \ref{fig:OhBo-chi} where the lighter yellow regions enlarge, while the darker zones shrink as one increases $\chi/R_0$ from $0.05$ to $0.15$. 

\bibliographystyle{prsty_withtitle}
\bibliography{holeySheets}